\documentclass{article}
\usepackage{spconf,amsmath,graphicx}
\usepackage{enumitem}
\usepackage{amsmath}
\usepackage{amsfonts}
\usepackage{amssymb}
\usepackage{xcolor}
\usepackage{subcaption}
\usepackage{algorithm,algorithmic}
\usepackage{relsize}
\usepackage[export]{adjustbox}
\usepackage{array,multirow,graphicx}
\usepackage{multirow}
\usepackage{graphicx}
\usepackage{lscape}
\usepackage{amsthm}
\usepackage{bbm}
\usepackage{graphicx}
\usepackage{booktabs}
\usepackage{url}

\usepackage{cite}

\theoremstyle{plain}
\newtheorem{theorem}{Theorem}[section]

\newtheorem{remark}[theorem]{Remark}

\theoremstyle{definition}

\DeclareMathOperator{\DP}{DP}

\DeclareMathOperator*{\argmin}{argmin}
\title{Sequential Diffusion-Guided Deep Image Prior For Medical Image Reconstruction}
\name{%
    Shijun Liang$^{*1}$\thanks{*~Equal contribution. This work was supported in part by the National Science Foundation (NSF) grants CCF-2212065 and CCF-2212066.}%
    \qquad Ismail Alkhouri$^{*2,3}$%
    \qquad Qing Qu$^{3}$%
    \qquad Rongrong Wang$^{2,4}$%
    \qquad Saiprasad Ravishankar$^{2,1}$%
}
\address{%
    $^{1}$\small{Department of Biomedical Engineering, Michigan State University, East Lansing, MI, USA
}  
\\%
$^{2}$\small{Department of Computational Mathematics, Science \& Engineering, Michigan State University, East Lansing, MI, USA
} 
\\%
$^{3}$\small{Department of Electrical \& Computer Engineering, University of Michigan, Ann Arbor, MI, USA  
}
\\%
$^{4}$\small{Department of Mathematics, Michigan State University, East Lansing, MI, USA
}
}
\begin{document}
%
\maketitle

\begin{abstract}

Deep learning (DL) methods have been extensively applied to various image recovery problems, including magnetic resonance imaging (MRI) and computed tomography (CT) reconstruction. Beyond supervised models, other approaches have been recently explored including two key recent schemes: Deep Image Prior (DIP) that is an unsupervised scan-adaptive method that leverages the network architecture as implicit regularization but can suffer from noise overfitting, and diffusion models (DMs), where the sampling procedure of a pre-trained generative model is modified to allow sampling from the measurement-conditioned distribution through approximations. In this paper, we propose combining DIP and DMs for MRI and CT reconstruction, motivated by (\textit{i}) the impact of the DIP network input and (\textit{ii}) the use of DMs as diffusion purifiers (DPs). Specifically, we propose a sequential procedure that iteratively optimizes the DIP network with a DM-refined adaptive input using a loss with data consistency and autoencoding terms. We term the approach Seq\textbf{u}ential \textbf{Di}ffusion-\textbf{G}uided \textbf{DIP} (uDiG-DIP). Our experimental results demonstrate that uDiG-DIP achieves superior reconstruction results compared to leading DM-based baselines and the original DIP for MRI and CT tasks.
\end{abstract}
%
\section{Introduction}
\label{sec:intro}

A variety of imaging modalities are used in clinical practice and disease diagnosis including
Magnetic Resonance Imaging (MRI) and Computed Tomography (CT)~\cite{cahyawijaya2023biomedical}. Accurate image formation is critical in MRI and CT and is formulated as inverse problems. Deep learning (DL) has recently been applied to solve large-scale image reconstruction problems~\cite{Ravishankar2018DeepReconstruction}. Among DL techniques, Deep 
Image Prior (DIP) \cite{ulyanov2018deep} is an unsupervised one-shot method that optimizes the parameters of a CNN (e.g. U-Net~\cite{ronneberger2015u}) using a data consistency loss between the network output (reconstruction) and measurements. However, DIP-based methods are prone to overfitting due to overparameterization of the networks~\cite{liu2019image}. \textcolor{black}{To mitigate noise ovefitting, previous studies considered different approaches such as early stopping (ES) \cite{wang2021early}, network pruning \cite{ghosh2024optimal}, and regularization \cite{liang2024analysis}. ES and pruning methods require computing running variances and image-specific binary mask, respectively.}

Diffusion models (DMs) \cite{ho2020denoising, song2020score} are generative models that sample from the \textit{approximated} distribution of a training set. When used to solve inverse problems, the sampling steps are modified to sample from the conditional distribution \cite{chung2022diffusion, chung2022improving, chung2022score, li2024decoupled}. However, these modifications are based on approximations that may not be very accurate and/or suitable for all tasks. DMs have also been used to mitigate the impact of worst-case perturbations \cite{nie2022diffusion, alkhouri2024diffusion}, reduce motion artifacts in MRI~\cite{oh2023annealed}, and image editing~\cite{mengsdedit}. Here, pre-trained DMs are employed as Diffusion Purifiers (DPs) where both the forward and reverse processes are used.

\vspace{0.2cm}

\noindent\textbf{Contributions:} \textcolor{black}{In this paper, we propose combining DIP with DMs employed as DPs to mitigate the noise over-fitting issue of DIP. In particular, we introduce seq\textbf{u}ential \textbf{Di}ffusion-\textbf{G}uided \textbf{DIP} (uDiG-DIP) motivated by the impact of the DIP network input and the capability of DMs, employed as DPs, to refine inputs, where we use diffusion-refined input-adaptive loss.} Our experiments demonstrate the effectiveness of uDiG-DIP on MRI and CT reconstruction, outperforming DIP and DM-based baselines on standard evaluation metrics. 


\textcolor{black}{To the best of our knowledge, this is the first paper to explore this combination since the work in \cite{chung2024deep}, where the authors use DIP to enhance the out-of-distribution adaptation of DM-based 3D reconstruction solvers in a meta-learning framework where fine-tuning the DM is needed. Our approach differs not only in the application (2D vs. 3D) and task (domain adaptation vs. reconstruction), but also in terms of methodology: we leverage pre-trained DMs as diffusion purifiers (DPs) within an input-adaptive DIP loss to reduce noise overfitting.}

\section{Preliminaries}
\label{sec:format}

Image reconstruction is an ill-posed inverse problem~\cite{compress} that seeks to recover an $n$-dimensional image $\mathbf{x}^*$ from an $m$-dimensional measurements vector $\mathbf{y}$, where $m<n$. The forward model can be formulated in different applications as $\mathbf{y}\approx\mathbf{A}\mathbf{x}^*$, where $\mathbf{A}$ is the forward operator. For multi-coil MRI, $\mathbf{A} = \mathbf{M} \mathbf{F} \mathbf{S}$, where $\mathbf{M}$ denotes coil-wise undersampling, $\mathbf{F}$ is the coil-by-coil Fourier transform, and $\mathbf{S}$ represents sensitivity encoding with multiple coils. For CT, we use a simplified forward operator to study the sparse-views setting: $\mathbf{A} = \mathbf{C} \mathbf{R}$, where $\mathbf{C}$ selects specific projection views or angles, and $\mathbf{R}$ is the radon transform~\cite{barrett1993radon} (corresponding to parallel beam CT).

\noindent \textbf{Deep Image Prior:} Deep image prior (DIP) was introduced by \cite{ulyanov2018deep}, showing that a U-Net generator network’s architecture alone can capture substantial low-level image statistics even without prior learning. Specifically, the DIP image reconstruction is obtained through:
\begin{equation}
\label{eqn: standard DIP}
    \hat{\theta} = \argmin_{\theta}  ~ \| \mathbf{A} f_{\theta}(\mathbf{z}) - \mathbf{y} \|_2^2\:,\;\;\;\; \hat{\mathbf{x}}=  f_{\hat{\theta}}(\mathbf{z})\:,
\end{equation}
where $\hat{\mathbf{x}}$ is the reconstructed image, and $\theta$ corresponds to the parameters of a network $f$. The input to the network, $\mathbf{z}$, is randomly selected and remains fixed during optimization. Although standard DIP performs well on many tasks, determining the optimal number of iterations is challenging, as the network may eventually fit noise in $\mathbf{y}$ or undesired images from the null space of $\mathbf{A}$.

\vspace{0.1cm}
\noindent \textbf{Diffusion Purification:} Given a pre-trained DM network $s_\phi(.,.)$ and an input image $\mathbf{x}$, diffusion purification (DP) involves adding perturbations for $M$ steps (also known as the process switching time \cite{alkhouri2024diffusion}), followed by the reverse sampling process to denoise the image \cite{nie2022diffusion, alkhouri2024diffusion, oh2023annealed}. The discretized forward and reverse processes are defined as follows \cite{song2020score}:
\begin{equation}
\label{eqn: DP forward}
    \mathbf{x}_M = \sqrt{\Bar{\alpha}_M}  \mathbf{x} + \sqrt{1-\Bar{\alpha}_M}\epsilon\:,
\end{equation}
\begin{equation}
\label{eqn: DP reverse}
    \mathbf{x}_{i-1} = \frac{1}{\sqrt{1-\beta}_i} \Big( \mathbf{x}_i +\beta_i s_\phi(\mathbf{x}_i,i) \Big) + \sqrt{{\beta}_i}  \epsilon_i \:,\\
\end{equation} 
where $i\in \{M,M-1, \dots, 1\}$, $\beta_i >0$ controls the speed of diffusion, $\Bar{\alpha}_M = \Pi^{M}_{i=1} (1-\beta_i)$, and $\epsilon, \epsilon_i \sim \mathcal{N}(\mathbf{0}, \mathbf{I})$. In this paper, we will denote this procedure by 
\begin{equation}
\label{eqn: DP}
    \mathbf{x}_0 = \DP_\phi(\mathbf{x}, M)\:.
\end{equation}
%


\section{Sequential Diffusion-Guided DIP}
\label{sec:uDiG-DIP}

\begin{figure}[t]
\centering
\includegraphics[width=8.0cm]{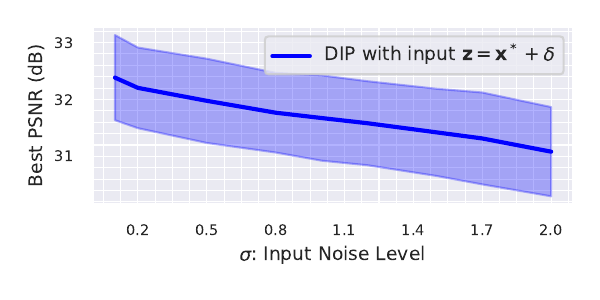}
\vspace{-0.7cm}
\caption{\small{Average best possible PSNR values (in dB) from standard DIP in \eqref{eqn: standard DIP} for 8 MRI scans (4x k-space undersampling) with respect to (w.r.t.) the perturbation level in $\mathbf{z}$.}}
\label{fig: impact of input}
\vspace{-0.7cm}
\end{figure}

\noindent \textbf{Motivation:} The motivation of this work is two fold: First is the impact of the DIP network input on the reconstruction quality. Second is the capability of DP to refine inputs as evidenced in \cite{nie2022diffusion, alkhouri2024diffusion, oh2023annealed}. Instead of random input, works like~\cite{zhao2020reference} and \cite{tachella2021neural} have considered how a structured input can impact DIP performance. Here, we explore: \textit{How does using a noisy version of the ground truth, retaining some structure, as the fixed input to the vanilla DIP objective in \eqref{eqn: standard DIP}, affect performance?} To investigate, we conduct the following experiment. Consider the MRI task $\mathbf{y} \approx \mathbf{A}\mathbf{x}^*$. Let the input to the standard DIP objective in \eqref{eqn: standard DIP} be $\mathbf{z} = \mathbf{x}^* + \boldsymbol{\delta}$, where $\boldsymbol{\delta} \sim \mathcal{N}(\mathbf{0}, \sigma^2 \mathbf{I})$. Here, $\sigma$ controls the noise level, with larger $\sigma$ yielding greater deviation between $\mathbf{z}$ and $\mathbf{x}^*$. We optimize \eqref{eqn: standard DIP} for different $\sigma$ values and record the highest PSNR before noise overfitting. Figure~\ref{fig: impact of input} shows that, for 8 images of the fastMRI dataset, the closer $\mathbf{z}$ to $\mathbf{x}^*$ (smaller $\sigma$) yields better reconstruction quality. This raises the question: \textit{Can we develop a diffusion-purified, input-adaptive DIP method to mitigate noise overfitting?} We address this question by proposing our method sequential Diffusion-Guided DIP (uDiG-DIP).

\vspace{0.2cm}

\noindent \textbf{uDiG-DIP:} Consider a (U-Net) neural network $f : \mathbb{R}^n \rightarrow \mathbb{R}^n$, parameterized by $\theta$, which takes input $\mathbf{z}$ and outputs $f_\theta (\mathbf{z})$\footnote{The entries of $\mathbf{x}^*$ and $\mathbf{y}$ in MRI are complex, so we use 2-channel networks for real and imaginary parts.}. Based on the insights about the DIP network input, we initialize $\mathbf{z}$ as $\mathbf{A}^H \mathbf{y}$. The initialization of $\theta$ follows standard DIP \cite{ulyanov2018deep}. Given the pre-trained DM, the proposed approach applies the following sequential procedure for $K$ iterations.
\begin{equation}
    \label{eqn: uDiGDIP param}
    \theta \leftarrow \argmin_{\theta} \|\mathbf{A} f_{\theta}(\mathbf{z}) -\mathbf{y}\|^2_2 + \lambda \|f_{\theta}(\mathbf{z}) -\mathbf{z} \|^2_2\:,
\end{equation}
\begin{equation}
    \label{eqn: uDiGDIP diffusion purification}
    \mathbf{z} \leftarrow \DP_\phi\big(f_{\theta}(\mathbf{z}),M\big) \:.
\end{equation}
In \eqref{eqn: uDiGDIP param}, $\lambda >0$ is a regularization parameter. The proposed procedure has two main components. First, it optimizes the network $f$ using an objective that includes a data consistency term and an autoencoding term designed to reduce noise overfitting. Second, it updates the input $\mathbf{z}$ after optimizing $\theta$, ensuring that the method remains \textit{diffusion-purified} and \textit{input-adaptive}.

\begin{figure}[t]
\centering
\includegraphics[width=8.5cm]{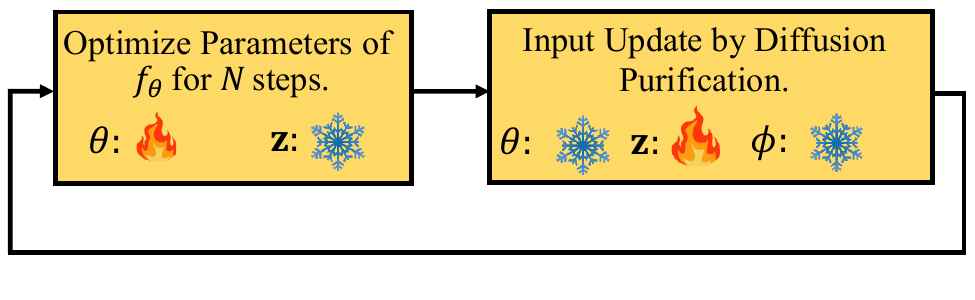}
\vspace{-0.5cm}
\caption{\small{A block diagram of the proposed uDiG-DIP procedure, run for $K$ sequential iterations.}} 
\label{fig: BD}
\vspace{-0.55cm}
\end{figure}

Algorithm~\ref{alg: uDiG-DIP} presents the procedure of uDiG-DIP. As inputs, the algorithm takes measurements $\mathbf{y}$, forward operator $\mathbf{A}$, number of sequential steps $K$, number of optimization steps $N$, number of diffusion steps $M$, the pre-trained DM with parameters $\phi$, regularization parameter $\lambda$, and the learning rate $\beta$. The parameters of $f$ are randomly initialized, \textcolor{black}{whereas $\mathbf{z}$ in the first step is set to the aliased image $\mathbf{A}^H\mathbf{y}$}. Then, for every iteration in $\{1,\dots,K\}$, the network is optimized for $N$ steps using a gradient-based optimizer, such as gradient descent (as shown in Algorithm~\ref{alg: uDiG-DIP}) or Adam \cite{kingma2014adam}. Past every $N$ gradient updates, the input is purified using DP (step 4). Figure~\ref{fig: BD} presents a block diagram of our proposed uDiG-DIP method.
\begin{algorithm}[t]
\small
\caption{Seq\textbf{u}ential \textbf{Di}ffusion-\textbf{G}uided \textbf{DIP} (uDiG-DIP).}
\textbf{Input}: Measurements $\mathbf{y}$, forward operator $\mathbf{A}$, number of iterations $K$, number of gradient updates $N$ per iteration, regularization parameter $\lambda$, learning rate $\beta$, pre-trained DM with parameters $\phi$, and number of diffusion steps $M$. \\
\vspace{1mm}
\textbf{Output}: Reconstructed image $\hat{\mathbf{x}}$. \\
\vspace{1mm}
\textbf{Initialization}: $\mathbf{z} \leftarrow \mathbf{A}^H \mathbf{y}$, $\theta \sim \mathcal{N}(\mathbf{0}, \mathbf{I})$. 
\\
\vspace{1mm}
\small{1:}  \textbf{For} $K$ iterations, \textbf{Do}  \\
\vspace{1mm} 
\small{2:} \hspace{1mm}  \textbf{For} $N$ iterations, \textbf{Do} (Network parameters update loop)  \\
\vspace{1mm}
\small{3:} \hspace{8mm}  $\theta \leftarrow \theta - \beta \nabla_\theta \Big[ \|\mathbf{A}f_{\theta}(\mathbf{z}) - \mathbf{y} \|_2^2 + \lambda \|f_{\theta}(\mathbf{z}) - \mathbf{z}\|_2^2\Big]$.\\
\vspace{1mm}
\small{4:} \hspace{1mm} $\mathbf{z} \leftarrow \DP_\phi\big(f_{\theta}(\mathbf{z}),M\big)$. (Input update by diffusion purification)  \\
\vspace{1mm}
\small{5:} \textbf{Reconstructed image:}  $\hat{\mathbf{x}} =  f_{\theta}(\mathbf{z})$ \\
\vspace{1mm}
\vspace{-3.5mm}
\label{alg: uDiG-DIP}
\end{algorithm}

\vspace{0.2cm}
\noindent \textbf{Intuition behind Diffusion Purification:}  DMs are trained on fully sampled clean images, with the iterative reverse sampling steps designed to sample from the learned distribution of these images. In the diffusion purification (DP) part of uDiG-DIP, when an intermediate image is provided, applying reverse sampling on a noisy version of that image should ideally yield a signal sampled from the clean distribution. However, because data consistency is not enforced during DP's reverse steps, we utilize the DIP network.


\vspace{0.2cm}
\noindent \textbf{Intuition behind the Auto-encoding term:} In DIP-based methods, 
overfitting happens as the network increasingly fits its output to the subsampled measurements, $\mathbf{y}$, during optimization. The exact iteration when PSNR decay begins is unpredictable and varies across tasks and even among images within the same task. In uDiG-DIP, when the output of network $f_{\theta}$ improves over the previous step, the autoencoder term implicitly enforces similarity between the input and output, delaying the onset of noise overfitting. This delay occurs because the method not only ensures measurement consistency but also enforces similarity between the input and output. 

\begin{remark}\em{
    Most DM-based solvers for inverse imaging problems modify the reverse sampling steps to enforce data consistency such as the diffusion posterior sampling work in~\cite{chung2022diffusion}. In contrast, uDiG-DIP decouples data consistency enforcement from the reverse sampling process by using DIP for a few iterations before applying diffusion purification, similar to the approaches in \cite{li2024decoupled,zhang2024improving}. Unlike these studies, which focus on natural image restoration, our method is applied to medical image reconstruction and uses DIP specifically for enforcing data consistency.}
\end{remark}


\section{Experiments}
\label{sec:exp}
\begin{table}[t]
\small
    \centering
    \resizebox{0.48\textwidth}{!}{\begin{tabular}{ccccc}
    \toprule
    \textbf{Task} & \textbf{Method} & \textbf{PSNR: 4x, \textcolor{blue}{8x}} & \textbf{SSIM: 4x, \textcolor{blue}{8x}} & \textbf{Time}  \\
    \midrule
    \multirow{4}{*}{MRI} & Score-MRI & \footnotesize$33.53\pm 0.01$ , \textcolor{blue}{$32.27 \pm 0.032$}  &  \footnotesize$0.84 \pm 0.002$, \textcolor{blue}{$0.818 \pm 0.002$}& \footnotesize$6.2$ \\
    &Ref-G DIP & \footnotesize$33.17\pm 0.02$ , \textcolor{blue}{$30.67 \pm 0.018$}  &  \footnotesize$0.0892 \pm 0.002$, \textcolor{blue}{$0.0872 \pm 0.002$} & \footnotesize$2.5$ \\
    &DIP & \footnotesize$30.21\pm 0.012$ , \textcolor{blue}{$28.75 \pm 0.021$}  &  \footnotesize$0.865 \pm 0.002$, \textcolor{blue}{$0.842 \pm 0.002$} & \footnotesize$1.5$ \\
    &\textbf{uDiG-DIP}  & \footnotesize$34.46\pm 0.02$ , \textcolor{blue}{$33.20 \pm 0.03$}  &  \footnotesize$0.0932 \pm 0.002$, \textcolor{blue}{$0.912 \pm 0.002$} & \footnotesize$3.5$ \\
    \bottomrule
    \bottomrule
     \textbf{Task} & \textbf{Method} & \textbf{PSNR: 18 views, \textcolor{blue}{30 views}} & \textbf{SSIM: 18 views, \textcolor{blue}{30 views}} & \textbf{Time}  \\
    \bottomrule
    \multirow{6}{*}{CT} & MCG & \footnotesize$35.70\pm 0.02$ , \textcolor{blue}{$33.60 \pm 0.02$}  &  \footnotesize$0.965 \pm 0.002$, \textcolor{blue}{$0.95 \pm 0.002$} & \footnotesize$6.4$ \\
    &FBP & \footnotesize$22.92\pm 0.02$ , \textcolor{blue}{$19.52 \pm 0.02$}  &  \footnotesize$0.75 \pm 0.002$, \textcolor{blue}{$0.67 \pm 0.002$} & \footnotesize$0.2$ \\
    &Ref-G DIP & \footnotesize$31.2\pm 0.02$ , \textcolor{blue}{$28.31 \pm 0.02$}  &  \footnotesize$0.892 \pm 0.002$, \textcolor{blue}{$0.842 \pm 0.001$} & \footnotesize$2.5$ \\
    &DIP & \footnotesize$26.41\pm 0.02$ , \textcolor{blue}{$24.5 \pm 0.01$}  &  \footnotesize$0.79 \pm 0.002$, \textcolor{blue}{$0.772 \pm 0.002$} & \footnotesize$1.5$ \\
    &\textbf{uDiG-DIP} & \footnotesize$36.70\pm 0.02$ , \textcolor{blue}{$33.96 \pm 0.02$}  &  \footnotesize$0.972 \pm 0.002$, \textcolor{blue}{$0.954 \pm 0.002$} & \footnotesize$3.2$ \\
    \bottomrule
    \end{tabular}}
    \vspace{-0.2cm}
    \caption{\small{Average PSNR, SSIM, and run-time results of uDiG-DIP against the selected baselines using 30 MRI test scans and 30 CT test scans (randomly sampled from the test set). The run-time (minutes) results are averaged over the two settings. values past $\pm$ correspond to the standard deviation. }}
    \label{tab: main results 1}
    \vspace{-0.55cm}
\end{table}
\begin{figure*}[!t]
\begin{tabular}[b]{ccccc}
    \textbf{Ground Truth}&
    \textbf{Score-MRI}&
    \textbf{Ref-G DIP}&
    \textbf{DIP}&
    \textbf{uDiG-DIP (Ours)}\\      \includegraphics[width=.17\linewidth,valign=t]{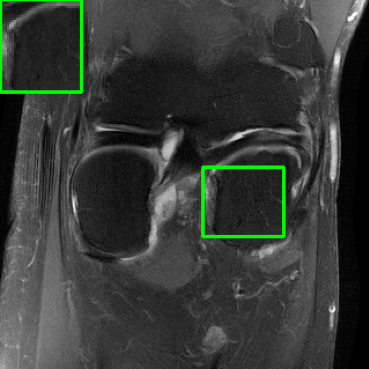}&
    \includegraphics[width=.17\linewidth,valign=t]{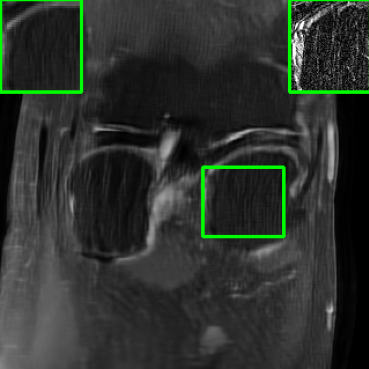}&
   \includegraphics[width=.17\linewidth,valign=t]{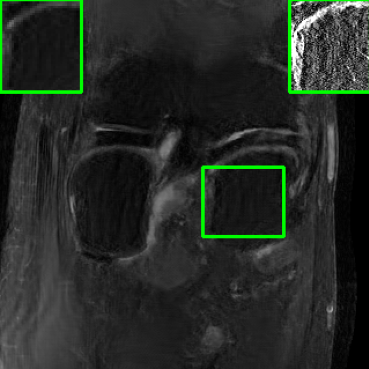} &
    \includegraphics[width=.17\linewidth,valign=t]{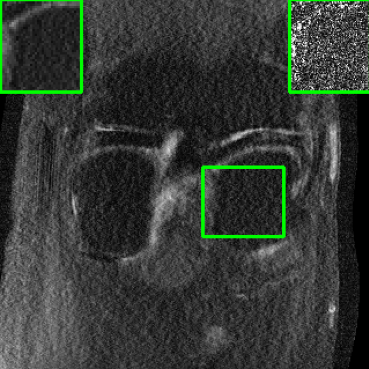}&
    \includegraphics[width=.17\linewidth,valign=t]{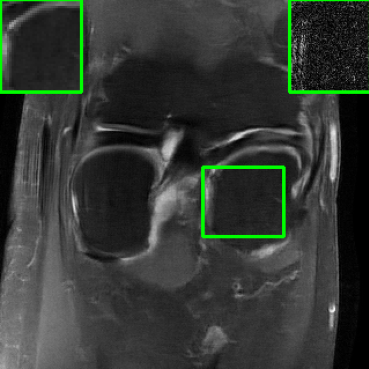}\\
    \scriptsize{PSNR = $\infty$  dB}  
   &\scriptsize{PSNR = 33.2 dB} 
    
    &\scriptsize{PSNR = 31.34 dB}
    &\scriptsize{PSNR = 30.23 dB } 
    &\scriptsize{\textbf{PSNR = 34.68 dB} }
        \\ 
    
\end{tabular}
\vspace{-0.45 cm}
\caption{\small{MRI visualizations of ground-truth and reconstructed images using different methods.}}
\label{fig: MRI_denoised_imgs_zoomed_300case}
\vspace{-0.18 in}
\end{figure*}

\begin{figure*}[!t]
\begin{tabular}[b]{ccccc}
    \textbf{Ground Truth}&
    \textbf{MCG}&
    \textbf{Ref-G DIP}&
    \textbf{DIP}&
    \textbf{uDiG-DIP (Ours)}\\      \includegraphics[width=.17\linewidth,valign=t]{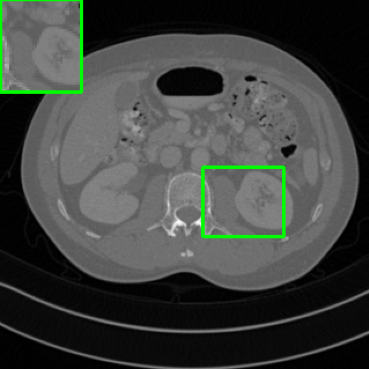}&
    \includegraphics[width=.17\linewidth,valign=t]{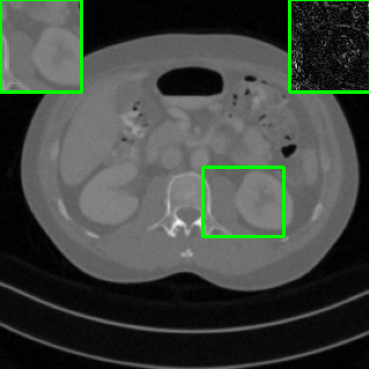}&
   \includegraphics[width=.17\linewidth,valign=t]{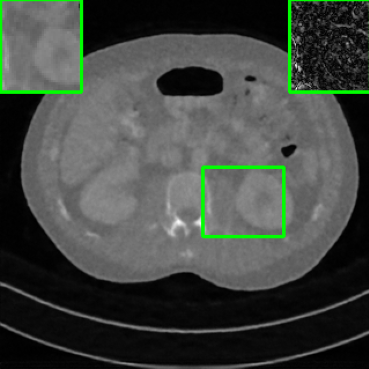} &
    \includegraphics[width=.17\linewidth,valign=t]{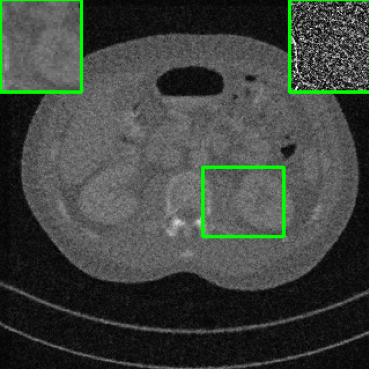}&
    \includegraphics[width=.17\linewidth,valign=t]{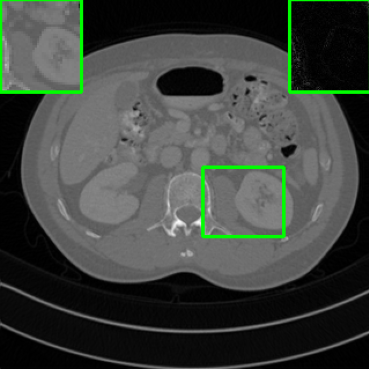}\\
    \scriptsize{PSNR = $\infty$  dB}  
   &\scriptsize{PSNR = 36.98 dB} 
    
    &\scriptsize{PSNR = 30.34 dB}
    &\scriptsize{PSNR = 22.23 dB } 
    &\scriptsize{\textbf{PSNR = 37.68 dB} }
        \\ 
    
\end{tabular}
\vspace{-0.45 cm}
\caption{\small{CT visualizations of ground-truth and reconstructed images using different methods.}}
\label{fig:CT_denoised_imgs_zoomed_300case}
\vspace{-0.25 in}
\end{figure*}

\begin{figure}[t]
\centering
\includegraphics[width=8.1cm]{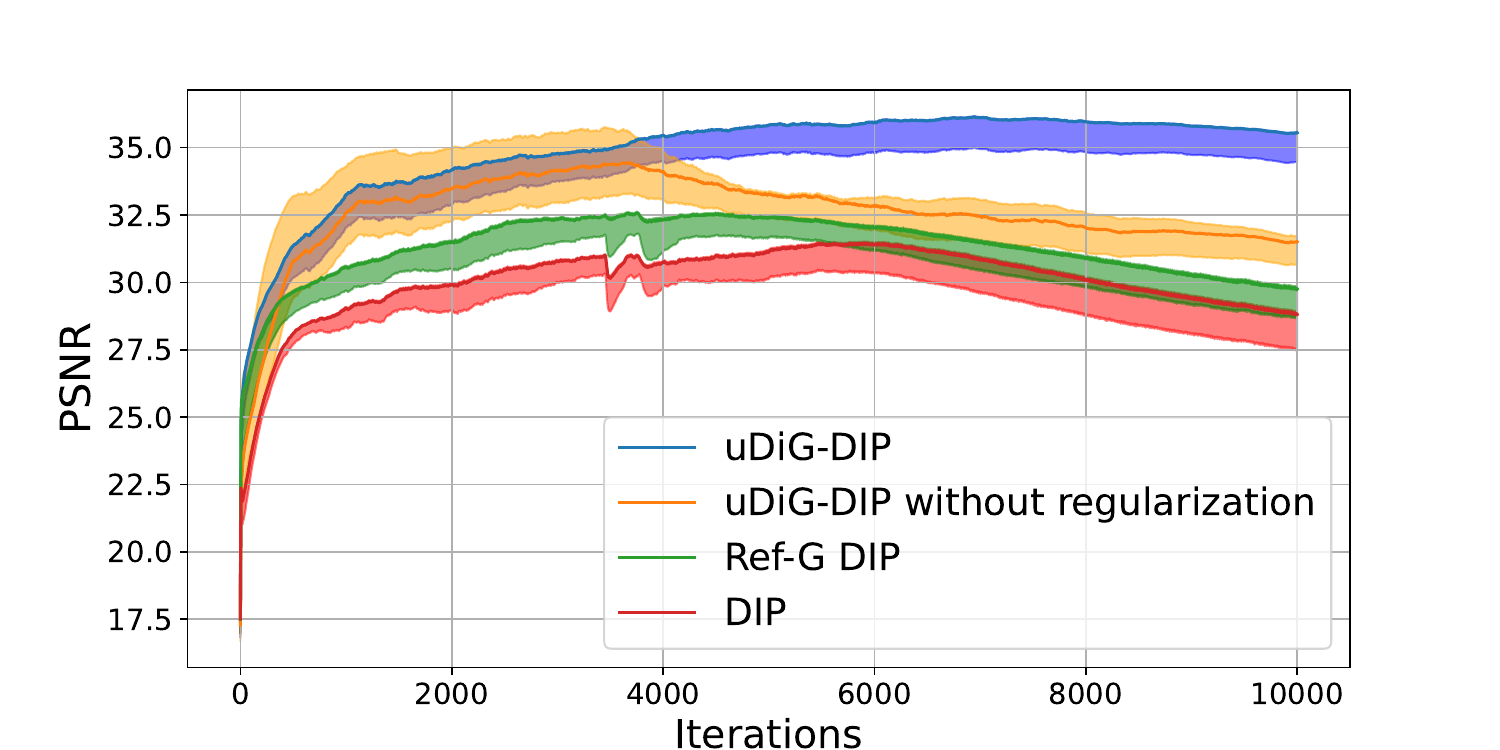}
\vspace{-0.4cm}
\caption{\small{Average PSNR values (in dB) w.r.t. overall iterations $\{1,\dots,NK\}$ for our method, and the network parameters optimization iterations for DIP and Ref-G DIP. Results are averaged over 30 MRI scans. uDiG-DIP without regularization corresponds to $\lambda = 0$ in \eqref{eqn: uDiGDIP param}.}}
\label{fig: overfitting performance}
\vspace{-0.8cm}
\end{figure}


\noindent \textbf{Experimental Setup:} In our experiments, we consider two medical imaging tasks: MRI reconstruction from undersampled measurements, and sparse-view CT image reconstruction. For MRI, we use the fastMRI dataset\footnote{\tiny{\url{https://github.com/microsoft/fastmri-plus/tree/main}}}. The multi-coil data is obtained using $15$ coils and is cropped to a resolution of $320 \times 320$ pixels. To simulate undersampling of the MRI k-space, we use a Cartesian mask with 4x and 8x accelerations. Sensitivity maps for the coils are obtained using the BART toolbox \cite{tamir2016generalized}. We use the parallel beam CT, we use the AAPM dataset\footnote{\tiny{\url{https://www.aapm.org/grandchallenge/lowdosect/}}}. The input image with $256 \times 256$ pixels is transformed into its sinogram representation using a Radon transform (the operator $\mathbf{A}$). The forward model assuming a monoenergetic source and no scatter for which the noise is $y_i = I_0 e^{-[\mathbf{A} \mathbf{x}^*]_i}$, with $I_0$ denoting the number of incident photons per ray (assumed to be $1$ for simplicity) and $i$ indexing the $i$th measurement or detector pixel. We use the post-log measurements for reconstruction, and a full set of $180$ projection angles to simulate two different sparse view acquisition scenarios (with equispaced angles). Specifically, we created cases with 18 and 30 angles/views. For the proposed uDiG-DIP method in Algorithm~\ref{alg: uDiG-DIP}, we use the Adam optimizer with learning rate of $\beta=0.0001$. The regularization parameter is $\lambda=1$. We set $N=2$, $K = 2000$, and $M=150$. We use the pre-trained DMs (and their noise schedule in \eqref{eqn: DP}) in \cite{chung2022score} and \cite{chung2022improving} for MRI and CT, respectively. Our code is available online\footnote{\tiny{\url{https://github.com/sjames40/uDIG_DIP/tree/main}}}. For DIP baselines, we use DIP \cite{ulyanov2018deep}, Ref-Guided DIP (Ref-G DIP) \cite{zhao2020reference}. For DM-based baselines, we use Score-MRI \cite{chung2022score} for MRI, and MCG \cite{chung2022improving} for CT. For DIP and Ref-G DIP, we use 4000 iterations. The quality of reconstructed images is measured using the peak signal-to-noise ratio (PSNR) and the structure similarity index measure (SSIM) \cite{wang2004image}.

\vspace{0.2cm}
\noindent \textbf{Main Results:} Table~\ref{tab: main results 1} presents the main quantitative results of uDiG-DIP and the considered baselines. As shown, uDiG-DIP achieves the best reconstruction results across both tasks and under the two settings of each task (Ax in MRI and views in CT). When compared to DM-based baselines (Score-MRI and MCG), we achieve slightly improved PSNR and SSIM scores while requiring significantly less runtime. This is due the requirement of Score-MRI and MCG to run a large number of sampling steps (500). When compared to DIP methods, uDiG-DIP achieves approximately 1 to 3 dB improvement for MRI and nearly 5 dB improvement for CT. Visualizations for 4x MRI and 30 views CT are provided in Figure~\ref{fig: MRI_denoised_imgs_zoomed_300case} and Figure~\ref{fig:CT_denoised_imgs_zoomed_300case}, respectively. The top-right green box shows the error in the region of the middle green box. As observed, uDiG-DIP achieves the best results, as also indicated by the PSNR values at the bottom of each reconstructed image.

\vspace{0.11cm}
\noindent \textbf{Robustness to Noise Overfitting Results:} In Figure~\ref{fig: overfitting performance}, we present the average PSNR values w.r.t. the overall optimization iterations of 30 MRI scans for uDiG-DIP, DIP, and Ref-G DIP. As observed, PSNR decay for DIP and Ref-G DIP begins at approximately iterations 6000 and 5000, respectively. For uDiG-DIP, PSNR decay starts around iteration 8000, indicating higher robustness to noise overfitting. This behavior is due to the use of the DP update in \eqref{eqn: uDiGDIP diffusion purification} and the auto-encoding regularization term in \eqref{eqn: uDiGDIP param}. To show the impact of the latter, we include the results of running uDiG-DIP without the regularization term, i.e., setting $\lambda$ to 0 (orange curve in Figure~\ref{fig: overfitting performance}). As observed, when $\lambda=0$, high PSNR values are maintained until about iteration 4000. However, compared to uDiG-DIP with $\lambda=1$, noise overfitting occurs earlier, demonstrating the effect of the auto-encoding regularization term.


\section{conclusion \& FUTURE WORK }
\label{sec:refs}

Motivated by the impact of the deep image prior (DIP) network input on reconstruction quality and leveraging diffusion models (DMs) as purifiers, we introduced a sequential method that alternates between optimizing the DIP network parameters using the standard data consistency term and an auto-encoding term, and updating the DIP network input using the pre-trained DM. We refer to this approach as sequential Diffusion-Guided DIP (uDiG-DIP). Our experimental results demonstrated the improved reconstruction quality of the proposed method in two medical imaging tasks, outperforming both DIP-based and DM-based baselines.

In future work, we plan to examine the effect of hyperparameters in uDiG-DIP, including the number of sampling steps in diffusion purification and the number of gradient updates. We also intend to explore the potential of this approach for natural image restoration tasks, both linear and nonlinear.

\small{
\bibliographystyle{IEEEbib}
\bibliography{refs}
}

\end{document}